\documentclass[conference]{IEEEtran}
\IEEEoverridecommandlockouts
\usepackage{cite}
\usepackage{amsmath,amssymb,amsfonts}
\usepackage{graphicx}
\usepackage{textcomp}
\usepackage{xcolor}
\def\BibTeX{{\rm B\kern-.05em{\sc i\kern-.025em b}\kern-.08em
    T\kern-.1667em\lower.7ex\hbox{E}\kern-.125emX}}
\begin{document}

\title{Frequency Components of Geomagnetically Induced Currents for Power System Modelling
\thanks{This work was supported in part by a grant from the Open Philanthropy Project and Eskom TESP funding.}
}


\author{\IEEEauthorblockN{David Oyedokun\IEEEauthorrefmark{1}, Michael Heyns\IEEEauthorrefmark{1}\IEEEauthorrefmark{2}, Pierre Cilliers\IEEEauthorrefmark{1}\IEEEauthorrefmark{2} and CT Gaunt\IEEEauthorrefmark{1}}
\IEEEauthorblockA{\IEEEauthorrefmark{1}Department of Electrical Engineering\\
University of Cape Town (UCT), Cape Town, South Africa}
\IEEEauthorblockA{\IEEEauthorrefmark{2}South African National Space Agency (SANSA), Hermanus, South Africa\\
davoyedokun@ieee.org}}

\maketitle

\begin{abstract}
Geomagnetically induced currents (GICs) in power systems are  conventionally modelled as direct currents, based on their commonly described quasi-DC nature. For more representative power system dynamic response modelling, it is necessary to model GICs using low-frequency AC. After analysing spectra of geomagnetic field measurements and GICs in several power systems, preliminary results indicate that the GIC spectra depend on the magnitude and frequency spectrum of the geomagnetic disturbance, the earth conductivity and the electrical characteristics of the network. In this paper, analysis based on measured GIC power spectra suggests that the dominant frequencies to use in GIC modelling and simulation are typically below 50 mHz. The associated error introduced by only using these frequencies is also quantified. The results have implications for modelling the geoelectric fields inducing GICs, dynamic models of power system response and the need for GICs and their associated driving fields to be measured at intervals not exceeding 10 s.
\end{abstract}


\section{Introduction}
From the first papers addressing GICs \cite{Albertson1970}, almost all current literature cites the quasi-DC nature of the induced currents.  The frequency range of this quasi-DC range has been identified as between 1 mHz and 1 Hz, where the driving geomagnetic disturbances (GMDs) rarely exceed the 1 Hz level \cite{Pulkkinen2003}. It has further been suggested that 1 min sampling ($\approx$16 mHz) suffices for accurate GIC modelling \cite{Pulkkinen2006}. Since this range is significantly below the typical power frequency, GIC modelling has generally assumed a DC circuit with the GIC flow simulated by rotating a benchmark driving geoelectric field across the network \cite{Lehtinen1985}. One of the largest concerns regarding GICs in power systems is the impact GICs have on reactive power and voltage stability. Most work focussing on the relationship between reactive power demand in transformers and GIC has similarly defined this relationship through models that assume the GIC is DC. Such an assumption results in the linear relationship between GIC magnitude and the increase in reactive power demand often reported \cite{Molinski2002}. It is however well understood that GICs are not truly DC, but rather low-frequency currents. It has then become imperative from a power systems perspective, to develop models which closely represent real GIC and the response of power systems to GICs. As a result, recent work investigating the dynamic response of power systems to GICs has required us to relax the DC assumption. Specifically, the power system response is modelled using low-frequency AC instead of DC. This approach of low-frequency AC driving more accurately represents GICs, which by nature are not strictly DC, and leads to significantly different responses.

When it comes to empirical analysis of the quasi-DC nature of GICs, less work has been done. Recently, NERC has established a GMD benchmark for GIC studies (TPL-007-1) using the geomagnetic (B) and geoelectric (E) field profiles for the Ottawa geomagnetic observatory during the 1989 GMD at 10 s intervals \cite{NERC}. Such sampling agrees with research showing 10 s cadence GIC data captures the GIC waveform without losing much information that could be useful in understanding the time response of the transformer and the power system \cite{Oyedokun2015}. From a GIC measurement perspective, EPRI's Sunburst project introduced 2 s cadence measurement of GICs during the 1990s which fulfils this requirement.  However, most utilities collecting GIC data often adopt longer measurement intervals (Australia records at 4 s and Eskom records at 5 min cadences respectively), believing them to be adequate. Looking at the geophysical drivers, until 10 years ago magnetic observatories typically only collected 1 min B-field data, with a few sites providing quasi-definitive 1 s cadence data, which makes further comparable analysis of the total GIC system difficult. Since then there has been a concerted move to 1 s cadence geomagnetic data, culminating in the new definitive 1 s cadence standard for INTERMAGNET that was adopted in 2014. The sampling rate of both the GICs and the driving B and E fields are critical when it comes to accurately describing the GICs in the frequency domain. The Nyquist criterion further requires sampling at least twice the frequency of components of interest - in this case the dominant frequency band for GIC modelling and simulation. Knowing the dominant frequency band of the GIC waveform is of particular interest in power system studies, in particular, voltage stability, as it relates to the true dynamic non-linear profile of the reactive power response of the transformer and the power system at large. Besides absolute magnitude, rapid changes in GIC of significant amplitude will lead to a different system response when compared to slowly varying GIC.

In this paper, the authors present an analysis based on measured GIC power spectra which was used to determine the characteristic frequency range for GIC modelling and simulation. The associated error introduced by using the frequency limit proposed is quantified.
\section{Theoretical Background}\label{sec:theory}
From first principles, GICs are driven by Faraday's law of induction. The changing flux of Earth's B-field through an induction loop, bounded by a grounded conductor at the surface (power network) and stretching deep into the conductive Earth, ultimately drives an electromotive force (EMF) that induces a current in the power system. The changing magnetic flux $dB/dt$, denoted as $Bdot$, is typically associated with a GMD arising from solar activity. In the frequency domain, the conductivity of the Earth relates the B-field to the E-field via the surface impedance $Z(f)$. The E-field in turn drives the EMF and, assuming the network is purely resistive, the GIC \cite{Lehtinen1985}. This assumption is a first order approximation, but in a real power system there are possible further higher order frequency dependent effects \cite{Weigel2019}.

Assuming a homogeneous Earth,
\begin{align}
E(f)=&Z(f)B(f), \label{eq:mt} \\
E(f)=&\sqrt{\dfrac{i2\pi f}{\mu \sigma}}B(f), \label{eq:mt_B}\\
\text{and }E(f)=&\dfrac{1}{\sqrt{\mu \sigma i 2 \pi f}}Bdot(f), \label{eq:mt_dB}
\end{align}
where $\sigma$ is the conductivity in S/m, $\mu$ is the permeability of Earth in H/m and $f$ the frequency in Hz.  The permeability of the Earth is taken as that of free space i.e. $\mu_0=4\pi\times 10^{-7}$ H/m. The conductivity of the Earth ranges from 1-100 mS/m. 

From \eqref{eq:mt_B}, it is evident that the Earth acts as a high-pass filter to generate the E-field from the B-field,
\begin{equation}
E(f)\propto \sqrt{f}B(f). \label{eqn:hpf}
\end{equation}
Given $Bdot$ in \eqref{eq:mt_dB}, the same Earth conductivity acts as a low-pass filter to generate the E-field,
\begin{equation}
E(f)\propto \frac{1}{\sqrt{f}}Bdot(f). \label{eqn:lpf}
\end{equation}
The B-field spectra has been shown to have `$1/f$' characteristics in its power spectrum. Given a power spectrum parametrisation of $f^{-m}$ (where $f$ is frequency), $m$ has been shown to be typically around 2, but can be as large as 4 \cite{Simpson2005}. The dominant periods in the B and E fields range from a couple of hours to tens of seconds, with a so called `dead-band' between 0.5 and 5 Hz \cite{Simpson2005,Jones1977}. It should be noted that a relationship proportional to $f^x$ in the equations above, would result in a relationship of $f^{2x}$ and slope of $2x$ in the power spectrum, since the power spectrum is proportional to magnitude squared. From this, and assuming the B-field power spectrum has a slope of $m$, the E-field power spectrum will have a less steep response, i.e. less than $m$. $Bdot$ on the other hand is proportional to $fB(f)$ and would have a power spectrum flatter than the B-field. In the case of $m\approx2$ for the B-field, the power spectrum of $Bdot$ would be flat, i.e. $m\approx0$. The E-field in turn is a low-pass filter of $Bdot$. Given a flat $Bdot$ response, the E-field would have $m\approx1$, which is consistent with $m<2$ as stated above. All this assumes a purely homogeneous Earth, which we know is not the case. The homogeneous Earth from a filter characteristic perspective is consistent nevertheless, even given particularly conductive or resistive profiles \cite{Zheng2013}. As mentioned above, the measured GIC includes the effect of the network itself which may introduce additional filtering effects, but this should not change the overall relative characteristics of the filters.
	
\begin{figure}[htbp]
\centerline{\includegraphics[width=0.5\textwidth]{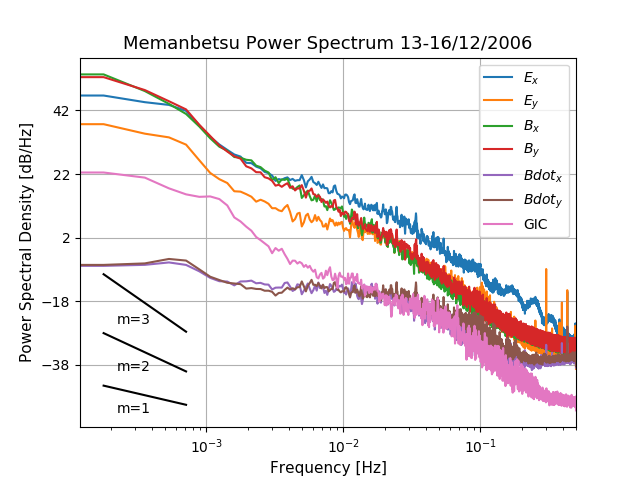}}
\caption{Comparison of different measured spectra response at MMB during a GMD. Various slopes as parametrised by $m$ are shown to guide the eye.}
\label{fig:mmb_all}
\end{figure}

At Memanbetsu (MMB), the Kakioka Magnetic Observatory of the Japan Meteorological Agency have 1 s cadence B-field and E-field measurements along with measured GIC as described by Watari \cite{Watari2009}. For a GMD, spectra of all the different components were computed (Fig. \ref{fig:mmb_all}) and found to be consistent with the theoretical relative filter characteristics described above, i.e. across almost all frequencies the GIC is a low-pass filter for $Bdot$, a high-pass filter for the B-field and similar to the E-field.

A measured E-field represents the most realistic version of \eqref{eq:mt}, where the surface impedance is fully 3D and takes into account all inhomogeneities. Far more typical is to use an assumed 1D layered-Earth model. This model typically averages out  inhomogeneities in the Earth over the network and in many cases is adequate for modelling, particularly when no surface impedance measurements are available \cite{Sun2019}. At MMB, there have been various studies around the $dB/dt$ proxy to GIC activity, with the B-field itself showing to be a better proxy \cite{Watari2009}. Depending on the conductivity profile, this may be the case when a poorly-conducting bottom layer underlies a thin good conductor \cite{Pirjola2010}. Regardless of profile, the lower frequencies dominate the GIC response, which is consistent with the quasi-DC assumption and theory.
\subsection{Skin Depth}
Using the concept of skin depth in conducting media, the cumulative spectral energy of a GIC signal can be used to indicate the typical depths and coarseness to which the Earth conductivity profile must be known in order to infer the relevant surface impedance needed in \eqref{eq:mt}.

The skin depth of an electromagnetic wave incident on a conducting medium is the depth at which the amplitude of the E-field is $1/e$ times the amplitude of the E-field at the surface of the conducting medium. For a homogeneous Earth as presented above, the skin depth is given by,
\begin{equation}
\delta=\sqrt{\dfrac{1}{\mu \sigma \pi f}}. \label{eq:skin}
\end{equation}

As a result, lower frequencies penetrate deeper into the Earth which increases the depth (and hence the cross sectional area) of the induction loop.
\begin{figure}[htbp]
\centerline{\includegraphics[width=0.5\textwidth]{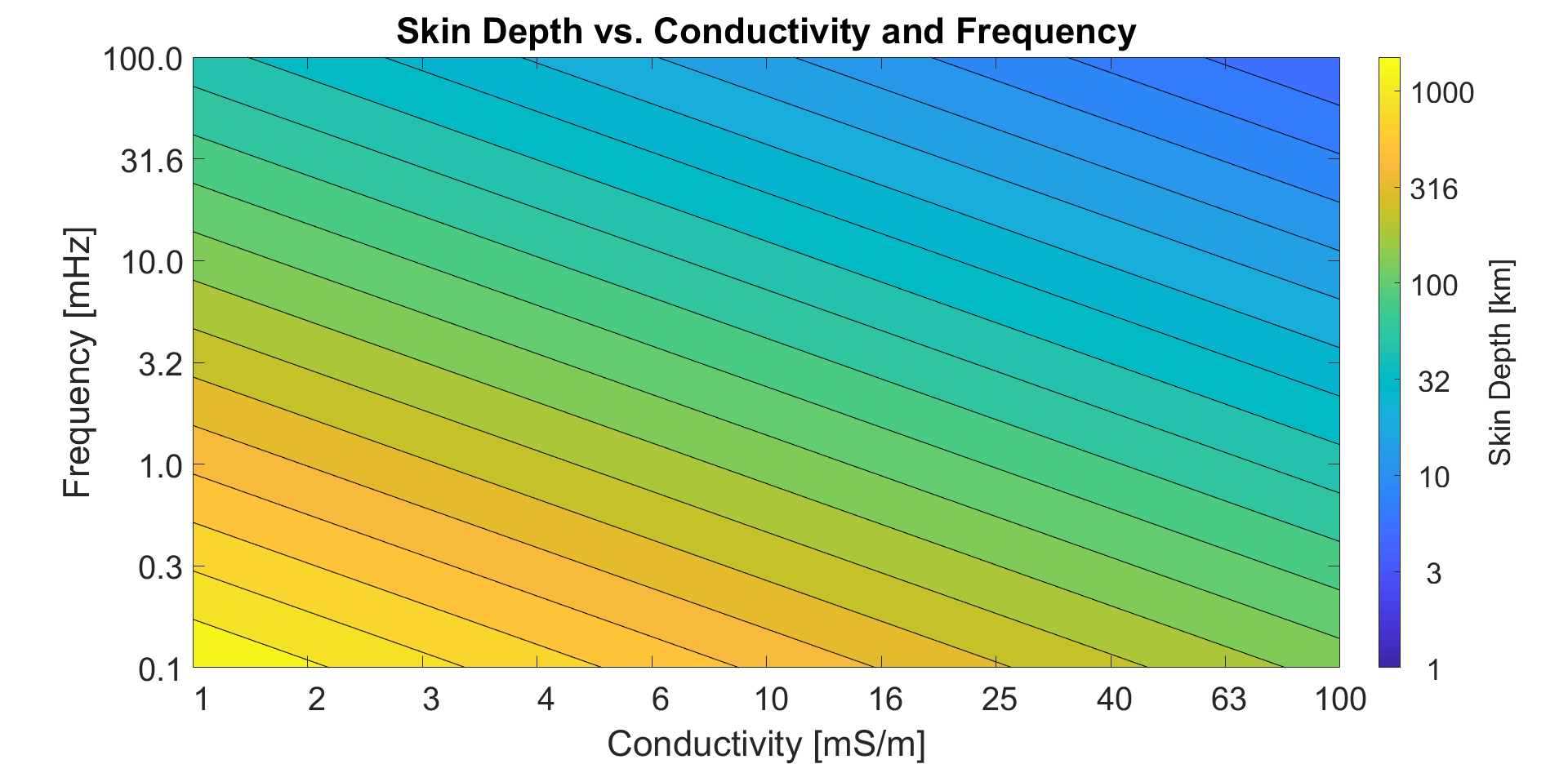}}
\caption{Skin depth (shown by a logarithmic colourbar) for a given frequency and conductivity assuming a homogeneous Earth. Given a frequency of 1 mHz and a conductivity of 1 mS/m, the skin depth is around 500 km.}
\label{fig:skin_depth}
\end{figure}
\section{Data Selection and Methodology}
In order to quantify the characteristic GIC frequencies, a number of GIC measurements from around the world were used for a 5 significant GMD events. These sites and GMDs are shown along with results in Tables \ref{tab1} and \ref{tab2}. The sites include GRS and HYD in South Africa, MONT and WEAK in the USA, MMB in Japan and LING in China. All these sites are at geomagnetic mid-latitudes except LING in China which is low-latitude. LING also has the added complication that it forms part of the Chinese HVDC network, which may introduce additional stray DC into the measured GIC \cite{Liu2009}. These sites and events were chosen for preliminary characterisation given a variety of conditions in regions that typically have significant power networks. Further research would look into characterising the spectral response of GICs in specific sub-regions and sub-events. 
\subsection{Raw Spectra}
To determine the spectra for each site over the different GMDs, all datasets were down-sampled to a common 2 s cadence. The power spectral density was then computed by Welch's average periodogram, using a 3 hour window. For GIC purposes, the 3 hour window captures the relevant peaks in a time-series, although there are lower frequencies that would be apparent such as daily variation and the 27 day solar rotation period. The 3 hour period is also indicative of distinct geomagnetic activity, and as such the well used geomagnetic K-index makes use of a 3 hour period. An example of such a raw spectrum is shown below in Fig. \ref{fig:grs_psd}.
\begin{figure}[htbp]
\centerline{\includegraphics[width=0.5\textwidth]{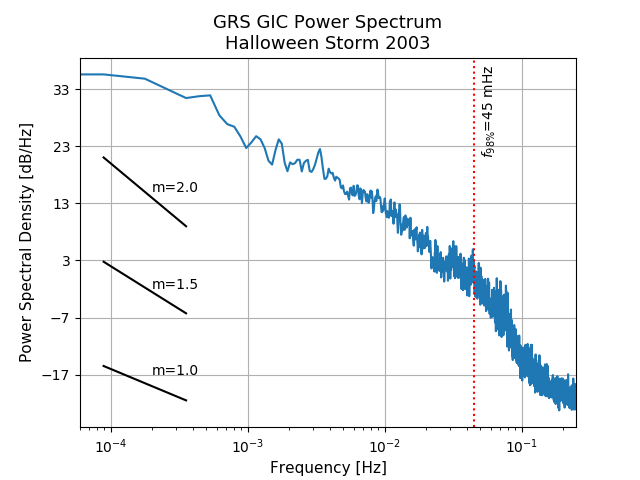}}
\caption{Typical power spectra of GRS GIC data during Halloween Storm of 2003. Various slopes as parametrised by $m$ in section \ref{sec:theory} are shown to guide the eye. 98\% of the spectrum energy sits to the left of the dashed red line.}
\label{fig:grs_psd}
\end{figure}
\subsection{Cumulative Spectra}
\begin{figure}[htbp]
\centerline{\includegraphics[width=0.5\textwidth]{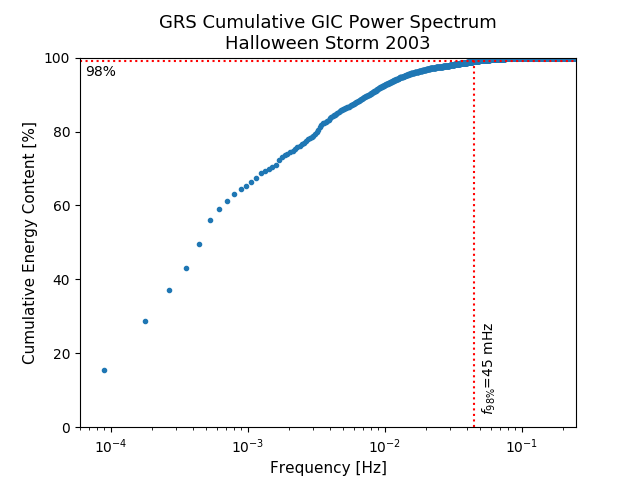}}
\caption{Cumulative power spectrum of GRS GIC. Also indicated is the ${f}_{98\%}$ frequency, under which 98\% of the cumulative power spectrum sits. If applied to a 1 ohm load, the cumulative power spectrum is equivalent to the cumulative energy content.}
\label{fig:grs_cumul}
\end{figure}

Using the raw spectra computed above, the cumulative energy in the signal can be computed. For this paper, the $f_{98\%}$ frequency is computed. As seen in Figs. \ref{fig:grs_psd} and Fig. \ref{fig:grs_cumul}, 98\% of the cumulative power spectrum or energy sits below $f_{98\%}$. Since all the datasets use the same approach to computing the spectra, the results are comparable. In the case of different sampling rates or window periods, $f_{98\%}$ may differ (see \ref{sec:dis}).
\subsection{Noise-level Spectra}
Another approach to computing the characteristic GIC spectra is to consider an extreme case of the spectrum during geomagnetically quiet periods. Practically this means the GIC amplitude during this period is very low with with no geomagnetic driving. The term `noise-level spectra' is used since compared to disturbed periods with significant GIC, these quiet periods would be classified as noise. The noise-spectra acts as an unbiased estimate of the system filter response, with the typical geomagnetic driving frequencies, such as those associated with pulsations, not complicating the spectra \cite{Boteler2001}. Considering the noise-spectra, there is not only the typical low-pass response with a negative slope in the range $-1$ to $-2$, but also a regime with a slope of zero at higher frequencies. A `knee' frequency $f_c$ splits the regimes. Any of frequencies above the $f_c$ could possibly be due to white noise which by definition has a flat spectrum. These white noise frequencies are further insignificant in comparison to the lower frequencies below the `knee'. During geomagnetic driving the spectrum is lifted out of the noise and the two regimes are not seen at all. 

\begin{figure}[htbp]
\centerline{\includegraphics[width=0.5\textwidth]{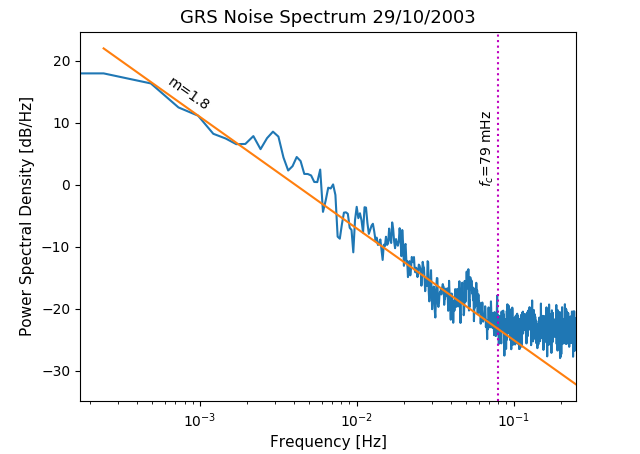}}
\caption{GIC noise-spectra at GRS prior to the Halloween Storm, with the associated power spectrum slope and `knee' frequency $f_c$ shown.}
\label{fig:hyd_noise}
\end{figure}

Since the noise power spectrum is typically more smooth than during a GMD, the slope of the power spectrum, such as in Fig. \ref{fig:hyd_noise}, gives an indication of the theoretical `$1/f$' weighting for frequencies. This weighting is critical to defining the frequency contributions in low-frequency modelling. During a GMD the slope remains in the form of `$1/f$', but may be less smooth and more difficult to fit. To generate these spectra, there has to be a feasible period of geomagnetically quiet time, which is not the case for the 2004 event. It should be noted that for this paper, $f_c$ was determined visually as the point where the power spectrum deviated from the fitted low-frequency power spectrum slope to a flat response. There would be some inherent uncertainty in this estimate as $f_c$ is not a single frequency, but rather a transition region.
\section{Results and Discussion}\label{sec:dis}
From the cumulative power spectrum analysis, 98\% of the energy spectrum of GICs is occurs at frequencies below 50 mHz, or periods more than 20 s. This in turn means that according to the Nyquist criterion a sampling cadence of better than 10 s is required to accurately represent the GIC signal in the frequency domain without significant aliasing, and likewise for the B-field and E-field. In terms of GIC modelling, the combination of frequency characteristics in Table \ref{tab1} and skin depth relation \eqref{eq:skin} allows one to estimate the depth dependence of the Earth's conductivity structure which determines the surface impedance, which is in turn needed for estimating the driving E-field from a measured B-field. For a ${f}_{98\%}$ of 50 mHz and conductivity of 1 mS/m, most of the power comes from frequencies penetrating to deeper than the skin depth of roughly 70 km. Changing the conductivity to 100 mS/m, the skin depth changes to roughly 7 km. At a 3 hour period and 1 mS/m, the skin depth stretches to roughly a depth of 1 600 km. These bounds suggest that the conductivity profile in an overly simplistic case without a significant geophysical strike (such as a coastline) does not have to be well resolved at shallow depths of a few kilometres to be representative of the majority of the GIC frequency content. These conclusions are drawn from a limited set of data which may not be representative of typical GICs. Further testing and comparison with high fidelity conductivity models needs to be done to determine whether coarse earth models are sufficient for GIC modelling on the basis that most of the current in the Earth flows at depths of hundreds of kilometres, where the conductivity profiles tend to be more homogeneous than closer to the surface.

\begin{table}[htbp]
\caption{Characteristics of measured GIC spectra for 5 GMD events}
\begin{center}
\begin{tabular}{|c|c|c|c|c|}
\hline
\rule{0pt}{2.3ex}  \textbf{GIC}&\textbf{GMD}&\multicolumn{3}{|c|}{\textbf{Frequency Response}} \\
\cline{3-5}
\rule{0pt}{2.3ex} \textbf{Site}&\textbf{Event}& $\mathbf{\textit{f}_{98\%}}$& \textbf{\textit{Noise Slope}}& $\mathbf{\textit{f}_{c}}$\\
\hline
\rule{0pt}{2.3ex} GRS & 29-31/10/2003 & 45 mHz & -1.8 & 79 mHz \\
\hline
\rule{0pt}{2.3ex} HYD & 07-10/11/2004 & 17 mHz & N/A & N/A \\
\hline
\rule{0pt}{2.3ex} LING & 07-10/11/2004 & 18 mHz & N/A & N/A \\
\hline
\rule{0pt}{2.3ex} HYD & 15-16/05/2005 & 35 mHz & -1.4 & 63 mHz \\
\hline
\rule{0pt}{2.3ex} MMB & 13-16/12/2006 & 2 mHz & -2.0 & 63 mHz \\
\hline
\rule{0pt}{2.3ex} MONT & 17-18/03/2015 & 25 mHz & -1.5 & 79 mHz \\
\hline
\rule{0pt}{2.3ex} WEAK & 17-18/03/2015 & 26 mHz & -1.3 & 79 mHz \\
\hline
\end{tabular}
\label{tab1}
\end{center}
\end{table}

In terms of the noise-spectra, we see that even at noise-levels, frequencies above 100 mHz carry no extra benefit. This is consistent with the tendency towards a dead-band of frequencies in the geomagnetic spectrum. Higher frequencies in the geoelectric spectrum are often associated with atmospherics which are not part of the process that drive GICs. The power spectrum slope is also constrained to between $-2$ and $-1$. This means that a theoretical frequency weighting relationship for GIC frequencies lies between $1/f$ and $1/\sqrt{f}$ and is applicable to weighting frequencies in GIC simulations.
\begin{figure}[htbp]
\centerline{\includegraphics[width=0.5\textwidth]{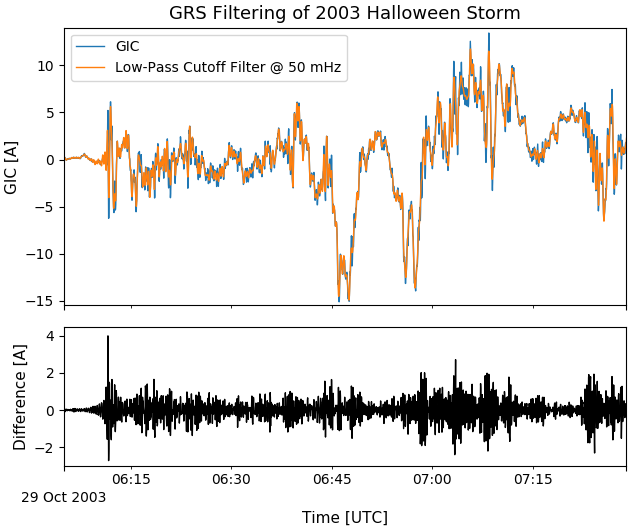}}
\caption{Time-series result of filter response and difference for GRS during the sudden storm commencement (SSC) and initial stages of the Halloween Storm. The maximum difference in this case is due to a shift in SSC peak by 4 s, which results in a 40\% difference when corrected for.}
\label{fig:mont_filter}
\end{figure}

A further consideration given the defined ${f}_{98\%}$ is what error is associated with ignoring higher frequencies. At this point it would be pertinent to note that the `true' ${f}_{98\%}$ cannot be determined by empirical analysis. In this paper we have a relative ${f}_{98\%}$ of the spectra between a period of 3 hours and twice the sampling cadence of 2 s. The `true' ${f}_{98\%}$ would need to take into account all frequencies. That said, since the tendencies of the driving fields are of a `$1/f$' nature, the effective ${f}_{98\%}$ used is a conservative estimate in the sense that there will be more power coming from frequencies lower than the 3 hour period than frequencies above the Nyquist frequency of the 2 s cadence data. In order to quantify the error introduced by filtering, the 2 s cadence raw data is used as a baseline. This corresponds to 500 mHz, which is also conservative considering 0.5 Hz is in the geomagnetic driving dead-band defined in theory \cite{Simpson2005} and above the `knee' frequency $f_c$ of the geomagnetically quiet time defined noise-spectra. To investigate the error associated with ignoring GIC frequency components above ${f}_{98\%}$, the 500 mHz measured GIC data were compared to the inverse Fourier transform of the corresponding GIC spectrum in which all parts of the spectrum above ${f}_{98\%}$ were set to zero. This is effectively a low-pass filter with a cutoff frequency at ${f}_{98\%}$. For each GIC dataset, the error resulting from the omission of higher frequencies is shown in Table \ref{tab2}. The various error metrics used include the standard RMSE and the maximum difference or error between the raw and low-pass filtered signals. These two metrics look at the typical error over the period and the worst-case error respectively. Also shown is the percentage error at the peak value or maximum of the GIC signal. This does not always correspond to the maximum error between the raw and low-pass filtered signals, but is of interest since mathematically peaks require higher frequencies to be resolved and peak GIC is often associated with the power system being at its most vulnerable during the height of the storm. The low-frequency contributions to these peak values nevertheless still dominate. The maximum error on the other hand is more often associated with the sudden storm commencement (SSC) at the start of the storm where there is a steep step due to the shock of the solar plasma slamming into Earth's B-field. Similarly, higher frequencies are needed to reproduce this type of sharp transient in the Fourier domain.
\begin{table}[htbp]
\caption{Characteristics of GIC errors due to low-pass filtering}
\begin{center}
\begin{tabular}{|c|c|c|c|c|c|}
\hline
\rule{0pt}{2.3ex}  \textbf{GIC}&\textbf{GMD}&\textbf{Max}&\multicolumn{3}{|c|}{\textbf{Error}} \\
\cline{4-6}
\rule{0pt}{2.3ex} \textbf{Site}&\textbf{Event}&\textbf{$|$GIC$|$}&\textbf{\textit{RMSE}} & \textbf{\textit{Max}} & \textbf{\textit{Peak (\%)$^{\mathrm{a}}$}} \\
\hline
\rule{0pt}{2.3ex} GRS & 29-31/10/2003 & 15.1 A & 0.1 A & 4.7 A & 12.7\% \\
\hline
\rule{0pt}{2.3ex} HYD & 07-10/11/2004 & 4.4 A & 0.1 A & 1.8 A & 2.0\% \\
\hline
\rule{0pt}{2.3ex} LING & 07-10/11/2004 & 72.9 A & 1.0 A & 13.0 A & 2.2\% \\
\hline
\rule{0pt}{2.3ex} HYD & 15-16/05/2005 & 5.7 A & 0.1 A & 1.9 A & 8.0\% \\
\hline
\rule{0pt}{2.3ex} MMB & 13-16/12/2006 & 4.0 A & 0.04 A & 0.9 A & 6.9\% \\
\hline
\rule{0pt}{2.3ex} MONT & 17-18/03/2015 & 14.3 A & 0.2 A & 4.4 A & 9.0\% \\
\hline
\rule{0pt}{2.3ex} WEAK & 17-18/03/2015 & 8.2 A & 0.1 A & 2.1 A & 0.6\%$^{\mathrm{b}}$ \\
\hline
\multicolumn{6}{r}{\scriptsize{$^{\mathrm{a}}$Percentage error in signal peak value, which is not always correlated to the max error}}\\
\multicolumn{6}{r}{\scriptsize{$^{\mathrm{b}}$Lesser peaks in main storm phase display peak error comparable to other sites}}
\end{tabular}
\label{tab2}
\end{center}
\end{table}

Although the low-pass filter results in Table \ref{tab2} seem adequate for GIC modelling, they do not tell the full story. If the GIC is under-sampled or low-pass filtered, the amplitude of the peaks may be underestimated. The extent of underestimation becomes worse with the decrease in the bandwidth of the GIC. Even if the bandwidth contains 98\% of the energy of the measured GIC, the underestimation of the peak amplitudes can be on the order of 10\% as seen in Table \ref{tab2}. The error in estimating the peak amplitude can increase significantly when there are steep step-like increases, with the accuracy of estimating the peak GIC during these periods severely limited. This effect is seen in passing the measured GRS GIC data through a low-pass filter with a cutoff frequency of 50 mHz (similar to ${f}_{98\%}$). The resulting time-series seen in Fig. \ref{fig:mont_filter} seems acceptable, with most of the profile shape accurately represented. However, the peaks around the time of the SSC, which is the part of the GIC with the fastest rise time but not in this case the largest peaks, are shifted and underestimated by up to 40\%. This is in line with the results of Grawe et al. \cite{Grawe2018} regarding the under-sampling of magnetic fields and its impact in the estimation of the peaks of the surface electric field. Even if the physical GIC drivers are not high-frequency in nature, mathematically these high frequencies are often needed to reproduce peaks (peak resolution) and sudden steep changes.
\section{Conclusion}
Given 2 s cadence GIC data and 3 hour windowing, it has been shown that at least 98\% of the energy bandwidth of the GIC is below 50 mHz. Comparing the often used 1 min resampling in GIC modelling due to B-field data cadence, a 16 mHz (1 min cadence) threshold in most cases includes between 90\% and 95\% of the GIC energy as defined by 2 s cadence and 3 hour windowing. Although the 50 mHz limit is not definitive, it allows for practical applications in GIC modelling and simulation. Limiting the bandwidth of the GIC to below 50 mHz  does however introduce an underestimation of the peak GIC on the order of 10\% and an underestimation of the GIC during the sudden storm commencement of up to 40\%, with varying degrees of underestimation given different sites and different GMDs.

In terms of simulating GIC using specific frequencies and weighting the contributions of each, the slope of the power spectrum can be used. Given all the datasets analysed, this slope sits between $m=1$ and $m=2$ or alternatively the relative frequency weighting in the GIC power spectrum is between $GIC(f)\propto 1/\sqrt{f}$ and $GIC(f)\propto 1/f$, which agrees with the underlying theory. The dominant frequency ranges also have implications in the estimation of the E-field used in traditional GIC source modelling. Due to the variation of skin depth with frequency, the accurate estimation of the surface impedance for the modelling of the low-end of the GIC spectrum requires a knowledge of the Earth conductivity for depths more than 10 km up to a couple of thousand kilometres. 

Since the sampling cadence of GIC data has been shown to significantly affect the waveform shape, it is important to use a cadence that describes both the geophysical drivers and power network effects. 50 mHz in this case seems reasonable and implies GIC data and geophysical data needs to be measured at a cadence of 10 s to reproduce the 50 mHz signal. A 10 s cadence is in line with the sampling interval used by NERC for benchmark GMD profiles to be used in GIC studies \cite{NERC}. Any GIC monitoring equipment should at very least be consistent with this threshold, but ideally further customisable.

The predominant frequency range and weighting factors defined in this paper are a necessary precursor to modelling the characteristics of GICs and understand the dynamic response of the power system, particularly the susceptibility of the power system to voltage instabilities. This is particularly important during moderate to severe GMD driving. Further studies may be able to refine these frequency thresholds.
\section*{Acknowledgment}
The authors acknowledge Eskom and the EPRI Sunburst project for measured GIC data in South Africa, Hokkaido\linebreak
\newpage
\noindent Electric Power in Japan and Tennessee Valley Authority in   the USA, along with S. Watari and C. M. Liu who facilitated the sharing of data from Japan and China respectively. Further thanks goes to S. I. Lotz at SANSA and R. S. Weigel from George Mason University for reviewing the manuscript and in particular for R. S. W.'s contributions regarding the underestimation of the GIC peaks due to filtering.


\vspace*{7pt}
\begin{thebibliography}{10}
\bibitem{Albertson1970} V. Albertson and J. Van Baelen, ``Electric and Magnetic Fields at the Earth's Surface Due to Auroral Currents'', \textit{IEEE Transactions on Power Apparatus and Systems}, PAS--89(4), pp. 578--584, 1970.
\bibitem{Pulkkinen2003} A. Pulkkinen, ``Geomagnetic Induction During Highly Disturbed Space Weather Conditions: Studies Of Ground Effects'', PhD Thesis, University of Helsinki, 2003.
\bibitem{Pulkkinen2006} A. Pulkkinen, A. Viljanen and R. J. Pirjola, ``Estimation of geomagnetically induced current levels from different input data'', \textit{Space Weather}, 4(8), pp. 1--15, 2006.
\bibitem{Lehtinen1985} M. Lehtinen and R. J. Pirjola, ``Currents produced in earthed conductor networks by geomagnetically-induced electric fields'', \textit{Annales Geophysicae}, 3(4), pp. 479--484, 1985.
\bibitem{Molinski2002} T. S. Molinski, ``Why utilities respect geomagnetically induced currents'', \textit{Journal of Atmospheric and Solar-Terrestrial Physics}, 64(16), pp. 1765–1778, 2002.
\bibitem{NERC} NERC TPL-007-1: Establish requirements for Transmission system planned performance during geomagnetic disturbance (GMD) events. North American Reliability Corp., 2017.
\bibitem{Oyedokun2015} D. T. O. Oyedokun, ``Geomagnetically Induced Currents (GIC) In Large Power Systems Including Transformer Time Response'', PhD Thesis, University of Cape Town, 2015.
\bibitem{Weigel2019} R. S. Weigel and P. J.  Cilliers, ``An Evaluation of the Frequency Independence Assumption of Power System Coefficients used in Geomagnetically Induced Current Estimates'', \textit{Space Weather}, 2019.
\bibitem{Simpson2005} F. Simpson and K. Bahr, Practical Magnetotellurics, Geophysical Journal International. Cambridge: Cambridge University Press, 2005.
\bibitem{Jones1977} A. G. Jones, ``Geomagnetic Induction Studies in Southern Scotland'', PhD Thesis, University of Edinburgh, 1977.
\bibitem{Zheng2013} K. Zheng et al., ``Effects of geophysical parameters on GIC illustrated by benchmark network modeling'', \textit{IEEE Transactions on Power Delivery}, 28(2), pp. 1183--1191, 2013.
\bibitem{Watari2009} S. Watari, ``Measurements of geomagnetically induced current in a power grid in Hokkaido, Japan'', \textit{Space Weather}, 7(3), pp. 1--11, 2009.
\bibitem{Sun2019} R. Sun and C. Balch, ``Comparison between 1-D and 3-D Geoelectric Field Methods to Calculate Geomagnetically Induced Currents: A Case Study'', \textit{IEEE Transactions on Power Delivery}, 2019.
\bibitem{Pirjola2010} R. J. Pirjola, ``Derivation of characteristics of the relation between geomagnetic and geoelectric variation fields from the surface impedance for a two-layer earth'', \textit{Earth, Planets and Space}, 62(3), 2010.
\bibitem{Liu2009} C. M. Liu, L. Liu and R. J. Pirjola, ``Geomagnetically Induced Currents in the High-Voltage Power Grid in China'', \textit{IEEE Transactions on Power Delivery}, 24(4), pp. 2368--2374, 2009.
\bibitem{Boteler2001} D. H. Boteler, ``Assessment of Geomagnetic Hazard to Power  Systems in Canada'', \textit{Natural Hazards}, 23(2/3), pp. 101--120, 2001.
\bibitem{Grawe2018} M. A. Grawe, J. J. Makela, M. D. Butala and F. Kamalabadi, ``The impact of magnetic field temporal sampling on modeled surface electric fields'', \textit{Space Weather}, 16(11), pp.1721--1739, 2018.
\end{thebibliography}
\end{document}